\documentclass[prl,aps,preprint,showpacs]{revtex4}
\usepackage{epsfig,amssymb,amsfonts}
\newcommand{\ket}[1]{|#1\rangle}
\newcommand{\bra}[1]{\langle #1|}

\newcommand{\AB}{\mbox{\tiny\rm AB}}
\newcommand{\A}{\mbox{\tiny\rm A}}
\newcommand{\B}{\mbox{\tiny\rm B}}

\begin{document}

\title{ Noiseless method for checking the Peres separability criterion by
local operations and classical communication}

\author{Yan-Kui Bai}
\affiliation{State Key Laboratory for Superlattices and
Microstructures, Institute of Semiconductors,\\ Chinese Academy of
Sciences, P. O. Box 912, Beijing 100083, P. R. China}
\author{Shu-Shen Li and Hou-Zhi Zheng}
\affiliation{ CCAST (World Lab.), P. O. Box 8730, Beijing 100080, P. R. China and\\
State Key Laboratory for Superlattices and Microstructures,
Institute of Semiconductors,\\ Chinese Academy of Sciences, P.O.
Box 912, Beijing 100083, P. R. China}

\begin{abstract}
We present a method for checking Peres separability criterion in
an arbitrary bipartite quantum state $\rho_{\AB}$ within local
operations and classical communication scenario. The method does
not require the prior state reconstruction and the structural
physical approximation. The main task for the two observers, Alice
and Bob, is to estimate some specific functions. After getting
these functions, they can determine the minimal eigenvalue of
$\rho^{T_{\B}}_{\AB}$, which serves as an entanglement indicator
in lower dimensions.
\end{abstract}


\maketitle

\section{I. introduction}
Quantum entanglement \cite{epr,sch,bel} has been an important
physical resource for quantum information processings \cite{nch},
for example, quantum teleportation, quantum key distribution, and
quantum dense code. Before we can make use of the entanglement, we
need know that it really exists in the system. The first and most
widely used criterion is the Peres separability criterion,
\emph{i.e.} the positive partial transpose (PPT) criterion
\cite{per,hhh}. If a quantum state $\rho_{\AB}$ has matrix
elements $\rho_{ij}^{mn}=\bra{ij}\rho_{\AB}\ket{mn}$ then the
partial transpose $\rho_{\AB}^{T_{\B}}$ is defined as
\begin{equation}\label{1}
    (\rho_{ij}^{mn})^{T_{\B}}=\rho_{in}^{mj}.
\end{equation}
The criterion is known if $\rho_{\AB}$ is separable, then it must
have a PPT. Thus any state for which $\rho_{\AB}^{T_{\B}}$ is not
positive semidefinite is necessarily entangled. When we deal with
an unknown quantum state, we can resort to quantum state
tomography \cite{vor} which provides the full knowledge about the
density matrix. However, there are more efficient ways that
compute the entangled properties directly via some functions of
density matrix $\rho_{\AB}$. A. Ekert and P. Horodecki \emph{et
al}. have done a series of works \cite{eke,pho,pha,phl,car} on
entanglement detection and measurement in an unknown mixed state
without the prior state reconstruction. These methods rely on two
techniques: the first is a modified interferometer network
\cite{eke} inserted a controlled-$U$ operation (for the analysis
c.f. \cite{sjo,fil}); the second is the structural physical
approximation (SPA) \cite{pho}, which achieve a non-physical map
approximately by mixing in an appropriate proportion the noise
operation $D(\rho)=I/d$. The SPA could tackle the problem of some
non-physical operation, but its practical implementation is
difficult. Recently, H. Carteret \cite{hac} constructed some
networks that can determine the eigenvalues of the partially
transposed density matrix $\rho_{\AB}^{T_{\B}}$, without resorting
to the SPA. This method is efficient and feasible for the physical
implementation.

In quantum communication, it is important to detect the
entanglement within local operations and classical communication
(LOCC) scenario, in which the two observers, Alice and Bob, are
far apart from each other and share a composite system. It has
been proven that entanglement is a precondition for secure quantum
key distribution \cite{mcu}. Based on the PPT criterion, C.M.
Alves \emph{et al}. presented a scheme to test the entanglement
with the aid of the LOCC implementation of the SPA \cite{car}. But
the physical implementation of the SPA is of more difficult in the
LOCC version.

In this paper, we present an LOCC method to check the Peres
separability criterion, an extension of H. Carteret's method
\cite{hac}. Our method is feasible for the physical implementation
in the LOCC scenario, because the SPA is not necessary. The main
task for Alice and Bob is to estimate some specific functions of
density matrix $\rho_{\AB}$ via two local networks. After getting
these functions, they can determine the spectrum of the matrix
$\rho_{\AB}^{T_{\B}}$ in which the minimal eigenvalue is an
entanglement witness.

This paper is organized as follows. In Sec. II, we present the
LOCC method for checking Peres separability criterion without SPA.
Then we discuss our method in Sec. III. Finally, in Sec. IV, we
give some conclusions.

\section{II. checking the PPT criterion by LOCC}
To see how the LOCC method works, we first recapitulate the global
method. In Ref. \cite{hac}, H. Carteret constructed some global
networks for estimating the eigenvalues of the partial transposed
matrix $\rho_{\AB}^{T_{\B}}$ without resorting to the SPA. In
general form, the network can be described by using Fig.1.
\begin{figure}[h]
\begin{center}
\epsfig{figure=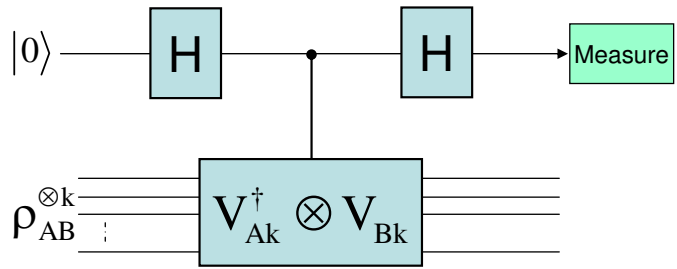,width=0.55\textwidth}
\end{center}
\caption{General form of Carteret's network, which can estimate
the eigenvalues of the partial transposed matrix
$\rho_{\AB}^{T_{\B}}$.}
\end{figure}
The method is partially inspired by the modified interferometer
network \cite{eke,sjo}, in which a controlled-$U$ operation is
inserted between two Hadamard gates. When one measures the control
qubit in the computational basis, the modification of interference
pattern is given by \cite{eke}
\begin{equation}\label{2}
    \mbox{Tr}(U\rho)=ve^{i\alpha},
\end{equation}
where $v$ is the visibility and $\alpha$ is the phase shift. H.
Carteret chooses the controlled-$U$ to be two controlled cyclic
permutations \cite{hac}, which is equivalent to the
controlled-$V_{Ak}^{\dagger}\otimes V_{Bk}$ as shown in Fig.1. The
unitary shift operator $V_{k}$ is defined as \cite{eke}
\begin{equation}\label{3}
    V_{k}\ket{\phi_{1}}\ket{\phi_{2}}\cdots\ket{\phi_{k}}
    =\ket{\phi_{k}}\ket{\phi_{1}}\cdots\ket{\phi_{k-1}},
\end{equation}
and $V_{Ak}^{\dagger}$ and $V_{Bk}$ act on the subsystems $A$ and
the subsystems $B$, respectively. By measuring the control qubit,
one can get the function $\mbox{Tr}[(V_{Ak}^{\dagger}\otimes
V_{Bk})\rho_{\AB}^{\otimes k}]$, which can be expanded into
\begin{eqnarray}\label{4}
    &&\mbox{Tr}[(V_{Ak}^{\dagger}\otimes V_{Bk})\rho_{\AB}^{\otimes k}]\nonumber\\
    &=&\mbox{Tr}\left[\sum
    \rho_{i_{1}j_{1}}^{m_{1}n_{1}}\rho_{i_{2}j_{2}}^{m_{2}n_{2}}\cdots
    \rho_{i_{k}j_{k}}^{m_{k}n_{k}}\ket{i_{1}j_{k}}\bra{m_{k}n_{1}}
    \otimes \ket{i_{2}j_{1}}\bra{m_{1}n_{2}}
    \otimes\cdots\otimes
    \ket{i_{k}j_{k-1}}\bra{m_{k-1}n_{k}}\right]\nonumber\\
    &=&\sum \rho_{i_{1}j_{1}}^{i_{2}j_{k}}\rho_{i_{2}j_{2}}^{i_{3}j_{1}}\cdots
    \rho_{i_{k}j_{k}}^{i_{1}j_{k-1}}.
\end{eqnarray}
Combining with Eq. (1), one can get the following relation
\begin{equation}\label{5}
    \mbox{Tr}[(V_{Ak}^{\dagger}\otimes V_{Bk})\rho_{\AB}^{\otimes
    k}]=\mbox{Tr}[(\rho_{\AB}^{T_{\B}})^{k}]=\sum_{i=1}^{d}\lambda_{i}^{k},
\end{equation}
in which $d$ denotes the dimension of $\rho_{\AB}$ and
$\lambda_{i}$ is the eigenvalue of $\rho_{\AB}^{T_{\B}}$. Thus, by
measuring $(d-1)$ functions, one can determine the spectrum of
$\rho_{\AB}^{T_{\B}}$.

In this paper, we present an LOCC method to check the Peres
separability criterion without resorting the SPA. It is assumed
that Alice and Bob share a number of the unknown quantum states
$\rho_{\AB}$. The main task of the two observers is to estimate
the function $\mbox{Tr}[(\rho_{\AB}^{T_{\B}})^{k}]$ within the
LOCC scenario. A normal LOCC network for the task is shown in
Fig.2.
\begin{figure}[h]
\begin{center}
\epsfig{figure=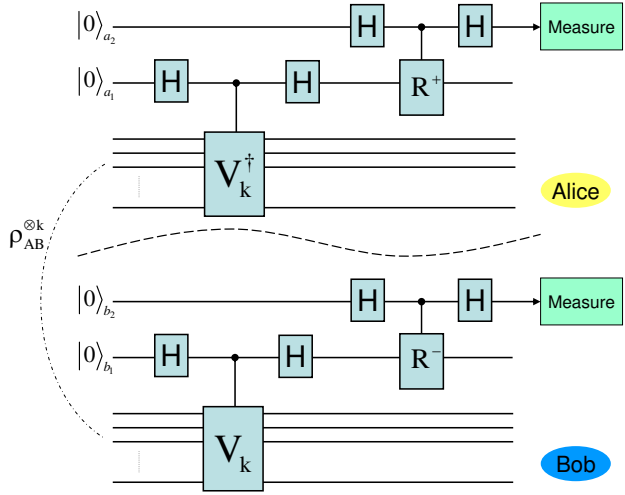,width=0.50\textwidth}
\end{center}
\caption{A normal network for remote estimation of the eigenvalues
of the partial transposed matrix $\rho_{\AB}^{T_{\B}}$. By
estimating the probabilities $P_{a_{2}b_{2}}(ij)$ that the two
ancillary qubits $a_{2}b_{2}$ is found in state $\ket{ij}$, Alice
and Bob can get the function
$\mbox{Tr}[(\rho_{\AB}^{T_{\B}})^{k}]$.}
\end{figure}
The network is composed of four modified interferometer circuits.
The first part for Alice is a modified interferometer circuit
which is attached to a controlled-$V_{Ak}^{\dagger}$ gate. The
ancillary qubit $a_{1}$ is the control qubit and the subsystems
$\rho_{\A}^{\otimes k}$ is the target. The second part following
the first is another interferometer circuit which is attached to a
controlled-$R^{+}$ gate. In this part, $a_{2}$ is the control
qubit and $a_{1}$ is the target qubit. The circuit for Bob is
similar to that for Alice except for some controlled quantum
gates. In the following analysis, we will show that Alice and Bob
can get the requisite function as long as they estimate the
probabilities $P_{a_{2}b_{2}}(ij)$ that in the measurement the two
ancillary qubits $a_{2}$ and $b_{2}$ are found in the state
$\ket{ij}$ where $i,j=0,1$.

We consider the first part of Alice and Bob's circuits. The input
state is
\begin{equation}\label{6}
    \rho_{in}(k)=\rho_{\AB}^{\otimes
    k}\otimes\rho_{a_{1}}\otimes\rho_{b_{1}},
\end{equation}
where $\rho_{a_{1}}=\ket{0}\bra{0}$ and
$\rho_{b_{1}}=\ket{0}\bra{0}$ are the initial states of the
ancillary qubits. The Hadamard gate and the controlled-$U$ gate in
their networks can be written as
\begin{eqnarray}\label{7}
&&H=\frac{1}{\sqrt{2}}\left(%
\begin{array}{cc}
  1 & 1 \\
  1 & -1 \\
\end{array}%
\right)\nonumber\\
&&U_{C-U}=\left(%
\begin{array}{cc}
  1 & 0 \\
  0 & 0 \\
\end{array}%
\right)\otimes
I+\left(%
\begin{array}{cc}
  0 & 0 \\
  0 & 1 \\
\end{array}%
\right)\otimes U
\end{eqnarray}
in the computational basis. After the two modified interferometer
circuits, the input state transforms into the following state
\begin{equation}\label{8}
    \rho'_{out}(k)=
U_{h}U_{v}U_{h}\rho_{in}(k)U_{h}^{\dag}U_{v}^{\dag}U_{h}^{\dag},
\end{equation}
where $U_{h}=H_{a_{1}}\otimes H_{b_{1}}\otimes I_{\AB}^{\otimes
k}$ and $U_{v}=U_{C-V_{A2k}^{\dagger}}\otimes U_{C-V_{B2k}}$. In
the state $\rho'_{out}(k)$, what we concern is the state evolution
of the two ancillary qubits $a_{1}$ and $b_{1}$. After some
deduction, we can obtain that the state of the two ancillary
qubits transforms into
\begin{eqnarray}\label{9}
\rho'_{a_{1}b_{1}}(k)&=&\mbox{Tr}_{AB}[\rho'_{out}(k)]\nonumber\\
&=&\frac{1}{4}\left(%
\begin{array}{cccc}
  1+\mu_{1}^{(k)}+\mu_{3}^{(k)} & \mu_{5}^{(k)} & \mu_{5}^{(k)} & -\mu_{4}^{(k)} \\
  -\mu_{5}^{(k)} & 1-\mu_{2}^{(k)}-\mu_{3}^{(k)} & \mu_{4}^{(k)} & -\mu_{5}^{(k)} \\
  -\mu_{5}^{(k)} & \mu_{4}^{(k)} & 1+\mu_{2}^{(k)}-\mu_{3}^{(k)} & -\mu_{5}^{(k)} \\
  -\mu_{4}^{(k)} & \mu_{5}^{(k)} & \mu_{5}^{(k)} & 1-\mu_{1}^{(k)}+\mu_{3}^{(k)} \\
\end{array}%
\right),
\end{eqnarray}
where
\begin{eqnarray}\label{10}
  \mu_{1}^{(k)} &=& \mbox{Tr}[(V_{Ak}\otimes I_{Bk})\rho_{\AB}^{\otimes k}]
  +\mbox{Tr}[(I_{Ak}\otimes V_{Bk})\rho_{\AB}^{\otimes k}], \nonumber \\
  \mu_{2}^{(k)} &=& \mbox{Tr}[(V_{Ak}\otimes I_{Bk})\rho_{\AB}^{\otimes k}]
  -\mbox{Tr}[(I_{Ak}\otimes V_{Bk})\rho_{\AB}^{\otimes k}], \nonumber \\
  \mu_{3}^{(k)} &=& \frac{1}{2}\mbox{Tr}[(V_{Ak}\otimes V_{Bk})\rho_{\AB}^{\otimes k}]
  +\frac{1}{4}\mbox{Tr}[(V_{Ak}^{\dag}\otimes V_{Bk})\rho_{\AB}^{\otimes k}]
  +\frac{1}{4}\mbox{Tr}[(V_{Ak}\otimes V_{Bk}^{\dag})\rho_{\AB}^{\otimes k}], \nonumber\\
  \mu_{4}^{(k)} &=& \frac{1}{2}\mbox{Tr}[(V_{Ak}\otimes V_{Bk})\rho_{\AB}^{\otimes k}]
  -\frac{1}{4}\mbox{Tr}[(V_{Ak}^{\dag}\otimes V_{Bk})\rho_{\AB}^{\otimes k}]
  -\frac{1}{4}\mbox{Tr}[(V_{Ak}\otimes V_{Bk}^{\dag})\rho_{\AB}^{\otimes k}], \nonumber\\
  \mu_{5}^{(k)} &=& \frac{1}{4}\mbox{Tr}[(V_{Ak}^{\dag}\otimes V_{Bk})\rho_{\AB}^{\otimes k}]
  -\frac{1}{4}\mbox{Tr}[(V_{Ak}\otimes V_{Bk}^{\dag})\rho_{\AB}^{\otimes k}].
\end{eqnarray}
Before considering the second part of the LOCC network, we need to
analyze Eq. (10) in detail. The shift operator $V_{k}$ has the
property \cite{phl}
\begin{equation}
\label{11} \mbox{Tr}(V_{k}\rho_{1}\otimes\rho_{2}\cdots\otimes
\rho_{k})=\mbox{Tr}(\rho_{1}\rho_{2}\cdots\rho_{k}).
\end{equation}
Based on the property, we can obtain $\mbox{Tr}[(V_{Ak}\otimes
I_{Bk})\rho_{\AB}^{\otimes k}]=\mbox{Tr}(\rho_{\A}^{k})$,
$\mbox{Tr}[(I_{Ak}\otimes V_{Bk})\rho_{\AB}^{\otimes
k}]=\mbox{Tr}(\rho_{\B}^{k})$ and $\mbox{Tr}[(V_{Ak}\otimes
V_{Bk})\rho_{\AB}^{\otimes k}]=\mbox{Tr}(\rho_{\AB}^{k})$. In Eq.
(5), we have $\mbox{Tr}[(V_{Ak}^{\dagger}\otimes
V_{Bk})\rho_{\AB}^{\otimes
k}]=\mbox{Tr}[(\rho_{\AB}^{T_{\B}})^{k}]$. The partial transposed
matrix $\rho_{\AB}^{T_{\B}}$ is an Hermitian matrix, therefore its
eigenvalues are real. Combining with the relation
$\mbox{Tr}(U^{\dagger}\rho)=[\mbox{Tr}(U\rho)]^{\ast}$, we can get
\begin{equation}\label{12}
    \mbox{Tr}[(V_{Ak}\otimes V_{Bk}^{\dag})\rho_{\AB}^{\otimes
    k}]=\mbox{Tr}[(V_{Ak}^{\dag}\otimes V_{Bk})\rho_{\AB}^{\otimes
    k}].
\end{equation}
Thus, in Eq. (10), the parameter $\mu_{5}^{(k)}$ equals zero. Now
the state of the ancillary qubits $a_{1}$ and $b_{1}$ can be
written in the following form
\begin{eqnarray}
\rho'_{a_{1}b_{1}}(k)
&=&\frac{1}{4}\left(%
\begin{array}{cccc}
  1+\mu_{1}^{(k)}+\mu_{3}^{(k)} & 0 & 0 & -\mu_{4}^{(k)} \\
  0 & 1-\mu_{2}^{(k)}-\mu_{3}^{(k)} & \mu_{4}^{(k)} & 0 \\
  0 & \mu_{4}^{(k)} & 1+\mu_{2}^{(k)}-\mu_{3}^{(k)} & 0 \\
  -\mu_{4}^{(k)} & 0 & 0 & 1-\mu_{1}^{(k)}+\mu_{3}^{(k)} \\
\end{array}%
\right),\nonumber
\end{eqnarray}
where
\begin{eqnarray}\label{13}
\mu_{1}^{(k)} &=& \mbox{Tr}(\rho_{\A}^{k})+\mbox{Tr}(\rho_{\B}^{k}),\nonumber\\
\mu_{2}^{(k)} &=& \mbox{Tr}(\rho_{\A}^{k})-\mbox{Tr}(\rho_{\B}^{k}),\nonumber\\
\mu_{3}^{(k)} &=&
\frac{1}{2}\mbox{Tr}(\rho_{\AB}^{k})+\frac{1}{2}\mbox{Tr}[(\rho_{\AB}^{T_{\B}})^{k}],\nonumber\\
\mu_{4}^{(k)} &=&
\frac{1}{2}\mbox{Tr}(\rho_{\AB}^{k})-\frac{1}{2}\mbox{Tr}[(\rho_{\AB}^{T_{\B}})^{k}].
\end{eqnarray}

In the second part of Alice and Bob's circuits, the input state is
$\rho'_{a_{1}b_{1}}(k)\otimes\rho_{a_{2}b_{2}}$, which is
subjected to two controlled operations $U_{C-R^{+}}$ and
$U_{C-R^{-}}$, where
\begin{eqnarray}\label{14}
&&R^{+}=\frac{1}{\sqrt{2}}(\sigma_{z}+\sigma_{y})=\frac{1}{\sqrt{2}}\left(%
\begin{array}{cc}
  1 & -i \\
  i & -1 \\
\end{array}%
\right),\nonumber\\
&&R^{-}=\frac{1}{\sqrt{2}}(\sigma_{z}-\sigma_{y})=\frac{1}{\sqrt{2}}\left(%
\begin{array}{cc}
  1 & i \\
  -i & -1 \\
\end{array}%
\right).
\end{eqnarray}
The initial state of the two control qubits $a_{2}$ and $b_{2}$ is
$\rho_{a_{2}b_{2}}=\ket{00}\bra{00}$. Beyond the second part, the
output state will be
\begin{equation}\label{15}
    \rho_{out}(k)=
U_{h}U_{r}U_{h}[\rho'_{a_{1}b_{1}}(k)\otimes
\rho_{a_{2}b_{2}}]U_{h}^{\dag}U_{r}^{\dag}U_{h}^{\dag},
\end{equation}
where $U_{r}=U_{C-R^{+}}\otimes U_{C-R^{-}}$. What we care about
is the evolution of the state $\rho_{a_{2}b_{2}}$. After some
deduction, we can obtain
\begin{eqnarray}\label{16}
\rho^{out}_{a_{2}b_{2}}(k)
&=&\frac{1}{4}\left(%
\begin{array}{cccc}
  1+\mu_{1}^{(k)}+\eta^{(k)} & 0 & 0 & 0 \\
  0 & 1-\mu_{2}^{(k)}-\eta^{(k)} & 0 & 0 \\
  0 & 0 & 1+\mu_{2}^{(k)}-\eta^{(k)} & 0 \\
  0 & 0 & 0 & 1-\mu_{1}^{(k)}+\eta^{(k)} \\
\end{array}%
\right),
\end{eqnarray}
where
$\eta^{(k)}=\mu_{3}^{(k)}-\mu_{4}^{(k)}=\mbox{Tr}[(\rho_{\AB}^{T_{\B}})^{k}]$.
With the aid of a classical communication, Alice and Bob can
estimate their probabilities $P_{a_{2}b_{2}}(ij)$ that in the
measurement the two qubits are found in the state
$\ket{ij}_{a_{2}b_{2}}$, here $i,j=0,1$. According to these
probabilities, they can get the function
$\mbox{Tr}[(\rho_{\AB}^{T_{\B}})^{k}]$, because
\begin{equation}\label{17}
 \eta^{(k)}=\mbox{Tr}[(\rho_{\AB}^{T_{\B}})^{k}]=P_{a_{2}b_{2}}(00)-P_{a_{2}b_{2}}(01)-P_{a_{2}b_{2}}(10)
 +P_{a_{2}b_{2}}(11).
\end{equation}
Therefore, for any $d_{\A}\otimes d_{\B}$ dimensional quantum
state $\rho_{\AB}$, Alice and Bob can determine the eigenvalues of
the partial transposed matrix $\rho_{\AB}^{T_{\B}}$ by estimating
the function $\mbox{Tr}[(\rho_{\AB}^{T_{\B}})^{k}]$ for
$k=2,3,\cdots,d_{\A}d_{\B}$. If the minimal eigenvalue
$\lambda_{min}$ is negative, the quantum state $\rho_{\AB}$ must
be entangled. This concludes our description of checking Peres
separability criterion within the LOCC scenario.

\section{III. discussions}
Among the functions $\mbox{Tr}[(\rho_{\AB}^{T_{\B}})^{k}]$,
$\mbox{Tr}[(\rho_{\AB}^{T_{\B}})^{2}]$ is a particular one. This
is because $V_{2}$ is the only Hermitian operator, compared with
the other shift operators $V_{k}$. According to Eq. (5), we have
\cite{hac}
\begin{equation}\label{18}
    \mbox{Tr}[(\rho_{\AB}^{T_{\B}})^{2}]=\mbox{Tr}[\rho_{\AB}^{2}].
\end{equation}
Inserting Eq. (18) in Eq. (13), we can see that the quantum state
$\rho'_{a_{1}b_{1}}(2)$ has the same form as that of
$\rho^{out}_{a_{2}b_{2}}(2)$. So, Alice and Bob can obtain the
eigenvalues of $\rho_{\AB}^{T_{\B}}$ by estimating the
probabilities $P_{a_{1}b_{1}}(ij)$. This means the second part of
the network is needless for estimating
$\mbox{Tr}[(\rho_{\AB}^{T_{\B}})^{2}]$. In this case, the LOCC
network shown in Fig.2 is the same as the network presented by
C.M. Alves \emph{et al} \cite{car}.

In Eq. (12), based on the Hermitian property of
$\rho_{\AB}^{T_{\B}}$, we have proved
$\mbox{Tr}[(V_{Ak}^{\dag}\otimes V_{Bk})\rho_{\AB}^{\otimes
k}]=\mbox{Tr}[(V_{Ak}\otimes V_{Bk}^{\dag})\rho_{\AB}^{\otimes
k}]$. The former function is
$\mbox{Tr}[(\rho_{\AB}^{T_{\B}})^{k}]$. Now we reanalyze the
latter function,
\begin{eqnarray}\label{19}
    &&\mbox{Tr}[(V_{Ak}\otimes V_{Bk}^{\dagger})\rho_{\AB}^{\otimes k}]\nonumber\\
    &=&\mbox{Tr}\left[\sum
    \rho_{i_{1}j_{1}}^{m_{1}n_{1}}\rho_{i_{2}j_{2}}^{m_{2}n_{2}}\cdots
    \rho_{i_{k}j_{k}}^{m_{k}n_{k}}\ket{i_{k}j_{1}}\bra{m_{1}n_{k}}
    \otimes \ket{i_{1}j_{2}}\bra{m_{2}n_{1}}
    \otimes\cdots\otimes
    \ket{i_{k-1}j_{k}}\bra{m_{k}n_{k-1}}\right]\nonumber\\
    &=&\sum \rho_{i_{1}j_{1}}^{i_{k}j_{2}}\rho_{i_{2}j_{2}}^{i_{1}j_{3}}\cdots
    \rho_{i_{k}j_{k}}^{i_{k-1}j_{1}}.
\end{eqnarray}
Having considered the definition of partial transposition, we can
get
\begin{equation}\label{20}
    \mbox{Tr}[(V_{Ak}\otimes V_{Bk}^{\dagger})\rho_{\AB}^{\otimes
    k}]=\mbox{Tr}[(\rho_{\AB}^{T_{\A}})^{k}].
\end{equation}
Therefore, in Fig.2, if Alice chooses the controlled-$V_{Ak}$ gate
and Bob chooses the controlled-$V_{Bk}^{\dagger}$ gate, they can
estimate the eigenvalues of $\rho_{\AB}^{T_{\A}}$. In fact,
$\rho_{\AB}^{T_{\B}}$ and $\rho_{\AB}^{T_{\A}}$ have the same
eigenvalues. Because, based on Eq. (12), we have
$\mbox{Tr}[(\rho_{\AB}^{T_{\B}})^{k}]=\mbox{Tr}[(\rho_{\AB}^{T_{\A}})^{k}]$.

Our LOCC method is more efficient compared with the LOCC quantum
state tomography. For an unknown two-qubit state, the LOCC quantum
tomography needs to estimate 15 parameters of
$\mbox{Tr}[(\sigma_{Ai}\otimes\sigma_{Bj})\rho_{\AB}]$ where
$\sigma_{i},\sigma_{j}=I_{2},\sigma_{x}, \sigma_{y},\sigma_{z}$.
However, our LOCC method needs to estimate only 3 parameters,
\emph{i.e.} $\mbox{Tr}[(\rho_{\AB}^{T_{\B}})^{2}]$,
$\mbox{Tr}[(\rho_{\AB}^{T_{\B}})^{3}]$ and
$\mbox{Tr}[(\rho_{\AB}^{T_{\B}})^{4}]$. In addition, compared with
the LOCC method  presented by C.M. Alves \emph{et al}. \cite{car},
our method is more feasible in the sense of physics since Alice
and Bob need not perform the SPA within the LOCC scenario.
Furthermore, the quantum network shown in Fig.2 is within the
reach of quantum technology currently developed.

For higher-dimensional bipartite systems, the Peres separability
criterion is only the necessary condition for entanglement
detection. There is a special type of quantum state--- bound
entangled state \cite{hla,ppp}, which has the property of PPT.
A.C. Doherty \emph{et al.} presented the notation of the PPT
symmetric extensions \cite{drl,doh}, which is a necessary and
sufficient condition for detecting bipartite entanglement. How to
efficiently check the PPT symmetric extension without the prior
state reconstruction is a considerable problem.

\section{IV. conclusions}
In this paper, we present a method for checking the Peres
separability criterion without resorting to the prior state
reconstruction and the SPA, which is an LOCC extension of H.
Carteret's method \cite{hac}. The LOCC method is more efficient
than the LOCC quantum state tomography. In addition, the method is
more feasible in the physical implementation than the LOCC method
presented by C.M. Alves \emph{et al} \cite{car}.

\section{acknowledgements}
This work was supported by the National Natural Science Foundation
of China (Grant Nos. 60325416 and 60328407) and the Special
Foundation for State Major Basic Research Program of China (Grant
No. G2001CB309500).

\end{document}